\documentclass{article}

\usepackage{PRIMEarxiv}

\usepackage[utf8]{inputenc} % allow utf-8 input
\usepackage[T1]{fontenc}    % use 8-bit T1 fonts
\usepackage{hyperref}       % hyperlinks
\usepackage{url}            % simple URL typesetting
\usepackage{booktabs}       % professional-quality tables
\usepackage{amsfonts}       % blackboard math symbols
\usepackage{nicefrac}       % compact symbols for 1/2, etc.
\usepackage{microtype}      % microtypography
\usepackage{lipsum}
\usepackage{fancyhdr}       % header
\usepackage{graphicx}       % graphics
\graphicspath{{media/}}     % organize your images and other figures under media/ folder

\usepackage{textcomp}
\usepackage{booktabs}
\usepackage{booktabs}
\usepackage{siunitx} 
\usepackage{array}   
\usepackage{multirow}
\usepackage{tabularx}
\usepackage{hyperref}

\title{CT Liver Segmentation via PVT-based Encoding and Refined Decoding}

\author{Debesh Jha, Nikhil Kumar Tomar, Koushik Biswas, Gorkem Durak, Alpay Medetalibeyoglu,\\
\textbf{{Matthew Antalek,  Yury Velichko, Daniela Ladner, Amir Borhani, Ulas Bagci}} \\
Machine \& Hybrid Intelligence Lab, Department of Radiology, Northwestern University \\
}

\begin{document}
\maketitle

\begin{abstract}
Accurate liver segmentation from CT scans is essential for effective diagnosis and treatment planning. Computer-aided diagnosis systems promise to improve the precision of liver disease diagnosis, disease progression, and treatment planning. In response to the need, we propose a novel deep learning approach, \textit{\textbf{PVTFormer}}, that is built upon a pretrained pyramid vision transformer (PVT v2) combined with advanced residual upsampling and decoder block. By integrating a refined feature channel approach with a hierarchical decoding strategy, PVTFormer generates high quality segmentation masks by enhancing semantic features. Rigorous evaluation of the proposed method on Liver Tumor Segmentation Benchmark (LiTS) 2017 demonstrates that our proposed architecture not only achieves a high dice coefficient of 86.78\%, mIoU of 78.46\%, but also obtains a low HD of 3.50. The results underscore PVTFormer's efficacy in setting a new benchmark for state-of-the-art liver segmentation methods. The source code of the proposed PVTFormer is available at \url{https://github.com/DebeshJha/PVTFormer}.
\end{abstract}

% % keywords can be removed
% \keywords{First keyword \and Second keyword \and More}

\section{Introduction}
The liver is the largest solid organ in the human body, crucial for metabolic functions and digestive processes. Globally, liver cancer is the third leading cause of cancer-related deaths, highlighting its significant impact on public health~\cite{arnold2020global}. The liver is also a common site for metastases from various abdominal cancers, such as colon, rectum, and pancreas, as well as distant cancers like breast and lung. Therefore, accurate segmentation of the liver is crucial for targeted therapies and surgical planning~\cite{bilic2023liver}. With advancements in medical imaging technologies such as computed tomography (CT) and magnetic resonance imaging (MRI), it is possible to visualize and segment the liver with precision, leading to more accurate diagnosis and treatment strategies~\cite {gotra2017liver}.

CT exams are widely available and cost-effective. They are essential for examining liver diseases and detecting liver tumors, making them a common choice for clinical trials and preoperative planning~\cite{oliva2004liver}.  However, manual liver segmentation is a challenging task. It is time-consuming, operator-dependent, and lacks reproducibility. The liver varies in size and shape within the population and is located in close proximity to many organs due to its central location and size~\cite{gotra2017liver}.  Therefore, computer-aided diagnosis methods are necessary to improve patient care in examining liver diseases and detecting tumors, utilizing CT’s high-resolution imaging and contrast sensitivity. Ongoing research and technological advances in liver segmentation hold the promise of making therapeutic treatments more precise and effective.

Recently, there have been numerous works on liver segmentation~\cite{christ2016automatic,khoshkhabar2023automatic,zheng2022automatic,rahman2022deep}. Zhang et al.~\cite{zheng2022automatic} proposed a deep learning model that was based on 3D convolution and convolutional long short-term memory (C-LSTM) for hepatocellular carcinoma (HCC) lesion segmentation. The proposed model utilized 4D data from dynamic contrast-enhanced (DCE) magnetic resonance imaging (MRI) images to segment liver tumors. Similarly, Rahman et al.~\cite{rahman2022deep} proposed a novel algorithm for automatically segmenting liver tumors from CT images using a hybrid ResUNet~\cite{zhang2018road} model. The model was built on a combination of ResNet and U-Net~\cite{ronneberger2015u} networks and involved processes such as preprocessing, image augmentation, feature extraction and selection, reason of interest selection, and passing it through the ResUNet architecture for ROI tumor segmentation.

Li et al.~\cite{li2018h} proposed a densely connected UNet framework (H-DenseUNet) for liver and tumor segmentation. The architecture comprises a 2-D DenseUNet focused on efficiently extracting intra-slice features with a 3D component designed for hierarchically aggregating volumetric contexts. The dual element for the architecture addressed the limitations of 2D convolutions that ignore the volumetric information and the limitation of 3D convolutions that suffer from high computational costs. Vorontsov et al.~\cite{vorontsov2018liver} proposed an architecture that consists of two fully convolutional networks,
connected in tandem and trained together end-to-end for liver and tumor segmentation. The first FCN (FCN 1) generates a liver segmentation mask, while the second (FCN 2) uses the latent output from the first and the original input as an additional input to precisely delineate liver lesions.

The above work shows that there has been a significant effort for efficient liver and tumor segmentation. The literature primarily focuses on 2D or 3D convolutional networks, which can lead to a trade-off between computational load and the ability to capture intricate features essential for precise liver segmentation. Our model harmonizes these aspects by utilizing PVT v2 as a backbone, enriching feature representation through residual learning and effectively maintaining crucial information without the computational expense associated with 3D convolutions. This enables the architecture to capture detailed semantic features with more significant precision than traditional CNN-based models, which might not fully utilize the hierarchical nature of image features or demand for the increased computational demand.

The main contributions of the work are as follows: 

\begin{itemize}
\item   \textbf{PVTFormer architecture:} We propose a novel encoder-decoder based architecture, \textit{PVTFormer},  that utilizes PVT v2 as a backbone.  This model takes advantage of residual learning for enhanced feature representation. By combining PVT v2, Up block, and Decoder block, we construct an end-to-end segmentation pipeline that optimizes computational resources while preserving essential information, significantly improving segmentation performance. 

\item \textbf{Hierarchical decoding strategy:} Our approach includes a novel hierarchical decoding strategy that incorporates specialized upscaling in the \textit{Up block} with effective multi-scale feature fusion in the \textit{Decoder}. This approach significantly enhances the network's ability to delineate detailed semantic features, which is vital for precise liver segmentation.

\item  \textbf{{Systematic evaluation}}: We have evaluated PVTFormer against eight existing state-of-the-art methods. PVTFormer obtained the highest dice coefficient of 86.78\%, mean IoU of 78.46\%, and a low H.D of 3.50. This underscores the architecture's superiority in terms of accuracy and reliability in segmenting healthy liver tissue.
    
\end{itemize}

\begin{figure}[!t]
    \centering
    \includegraphics[width = 0.7\textwidth]{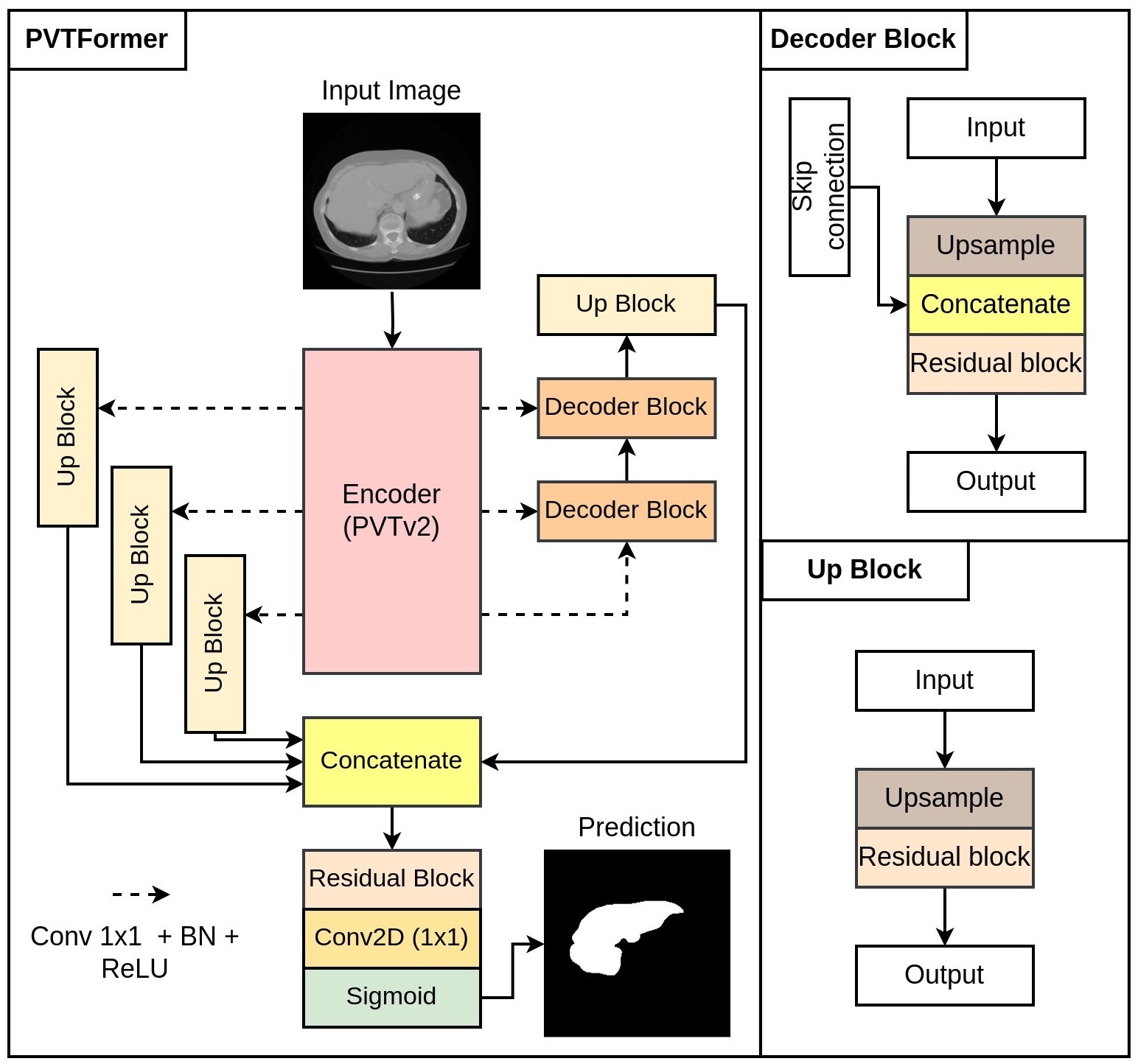}
\caption{Overview of the proposed \textit{PVTFormer} architecture: Input images are passed via PVT v2 encoder, generating three diverse feature maps which are passed through a series of ($1\times1$ Conv, BN, and ReLU)  before entering the Up block and the decoder block. The \textit{Up block} is passed through bilinear upsampling to upscale feature map dimensions to that of the input image. Next, a residual block fine-tunes the upsampled feature map, enhancing its ability to learn better representation.}
    \label{fig:PVTFormer}
\end{figure}
\section{Methodology}

\subsection{Encoder:}
Figure~\ref{fig:PVTFormer} shows the block diagram of the \textit{PVTFormer} architecture. Our architecture is an encoder-decoder-based framework and utilizes Pyramid Vision Transformer (PVT v2 b3)~\cite{wang2022pvt} as the pre-trained encoder. The main motivation behind adopting PVT v2 is to reduce the computational complexity and significantly improve liver segmentation tasks. The distinctive design of PVT v2 incorporates linear spatial reduction attention, overlapping patch embeddings, and a convolutional feed-forward network. The PVT v2 enhances the local continuity of images and feature map coherence while effectively handling variable-resolution input images with linear computational complexity compared to CNNs~\cite{wang2022pvt}. The input images are passed through the PVT v2 encoder, which captures hierarchical image features through overlapped patch embeddings and successive hierarchical transformer blocks, producing multi-scale representations. The encoder transforms input images into feature-rich representations and extracts three diverse feature maps at varying resolutions of $64\times64$,  $32\times32$, and  $16\times16$, having 64, 128, 320 as the number of feature channels respectively. These representations undergo a sequential transformation involving {$1\times1$ Convolution, Batch Normalization (BN), and Rectified Linear Unit (ReLU)} activation operations. This effectively reduces feature channels to $64$. This reduction optimizes computational efficiency and ensures the preservation of essential anatomical information crucial for precise liver segmentation.
 
% The linear spatial reduction attention reduces the high computational cost caused by attention operation. The overlapping patch embedding models the local continuity information and convolutional feed-forward network to enhance positional information utilization and spatial feature extraction, potentially improving spatial relationships and overall feature representation. 

\subsection{Up block:}
Figure~\ref{fig:PVTFormer} (right (bottom)) shows the block diagram of the \textit{Up block}.
The \textit{Up block} acts as a scaling unit to increase the spatial dimensions of feature maps. It comprises an \textit{upsampling layer} followed by a \textit{residual block}. Within the \textit{Up block}, the input feature map is first passed through a bilinear upsampling to upscale the feature map's height and width to that of the original input image. The \textit{residual block}, which consists of two convolutional operations with an identity mapping, refines the upscaled features, enabling the network to learn a more robust representation.

\subsection{Decoder block and output generation:}
The \textit{Decoder block} ({Figure~\ref{fig:PVTFormer} }(right (top)) fuses multi-scale features using \textit{skip connections} and \textit{upsampling layer}. It first upsamples the lower-resolution feature map and then concatenates it with a corresponding higher-resolution feature map from the encoder. This concatenated feature map is passed through another residual block to refine the combined features and enhance the segmentation accuracy for the liver structures. To produce the final segmentation output, the feature maps from both the \textit{Up blocks} and \textit{Decoder blocks} are \textit{concatenated} to integrate multi-scale contextual information. We apply a \textit{final residual block} to this integrated feature map to ensure uniform scale refinement. The refined feature map is then passed through a $1\times1$ Conv2D layer followed by a \textit{sigmoid activation} to generate the binary segmentation map.

\section{Experiments}
\label{sec:experiments}

\subsection{Datasets}
We utilize the Liver Tumor Segmentation Benchmark (LiTS)~\cite{bilic2023liver} dataset for our liver segmentation task. LiTs is a multi-center dataset collected from seven clinical centers from Germany, Netherlands, Canada, France and three medical centers from Israel. The dataset is entirely anonymous, and the personal identifier has been removed. It contains 201 CT images of the abdomen, of which 130 CT scans are made publicly available along with ground truth. Only the training dataset is made publically available. Thus, we divide the training dataset into three parts for our experimentation. 

To avoid bias, we split the cases into independent training (70 patients), validation (30 patients), and test (30 patients) sets. We resized the image to $256\times256$ pixels to optimize the trade-off between training time and model complexity. The volumetric CT scans were processed slice-by-slice to fit into regular computer hardware (GPU). During preprocessing, we extracted the healthy liver masks. Thus, we have 11684 slices in training dataset, 2745 in validation and 4734 slices in the testing set. The difference in slices in the validation and test is because the number of slices might vary between different patients.

\subsection{Implementation details}
We have used the PyTorch~\cite{paszke2019pytorch} framework for all the experiments used for liver segmentation tasks. All the experiments were performed on NVIDIA RTX A6000 GPU. The network is configured to train with a batch size 16 and a learning rate set to 1e$^{-4}$. We train all the models for 500 epochs to fine-tune the network parameters adequately with an early stopping patience of 50. To enhance the performance of our network,  we have used a combination of binary cross-entropy and dice loss and an Adam optimizer was chosen for parameter updates.

\subsection{Evaluation metrics}
Liver segmentation performance is evaluated using standard evaluation metrics such as dice coefficient (DSC), mean intersection over union (mIoU), recall, precision, F2 score and Hausdorff Distance (HD). Overlap-based metrics such as DSC and mIoU provide insight into the accuracy and reliability of overlap, whereas HD evaluates the differences between the segmented boundary and the actual liver boundary.

\begin{table*}[t!]
\centering
\caption{Model performance on the Liver Tumor Segmentation benchmark dataset~\cite{bilic2023liver}.}  
 \begin{tabular} {c|c|c|c|c|c|c|c}
\toprule

\textbf{{Method}}  & \textbf{{Publication}}  &\textbf{Dice (\%)} & \textbf{mIoU (\%)}  &\textbf{Recall (\%)} & \textbf{Precision (\%)} & \textbf{F2 (\%)} & \textbf{{HD}} \\ 
\hline

U-Net~\cite{jha2020doubleu}  &MICCAI 2015 & 82.06  & 73.40 & 77.82 & 91.10 &	78.98 & 3.79 \\

ResUNet++~\cite{jha2019resunet++}  & IEEE ISM 2019 & 77.03 & 68.19 & 79.90 & 83.57  & 76.24 & 3.96 \\

DoubleU-Net~\cite{jha2021real}  &IEEE CBMS 2020 & 86.24 & 77.89 & 80.68 & 95.00 & 81.28	& 3.69 \\

ColonSegNet~\cite{jha2021real}  & IEEE Access 2021 &80.87 &71.71  &80.07& 87.50 &79.22 & 3.84\\

NanoNet~\cite{jha2021nanonet}  &IEEE CBMS 2021 & 75.05 & 66.53 & 73.33 & 83.25 & 73.23 &4.01 \\

UNeXt~\cite{valanarasu2022unext} &MICCAI 2022	& 81.31 & 72.43 &80.73 & 87.40& 79.93 &3.74 \\

TransNetR~\cite{jha2023transnetr} & MIDL 2023 &86.11  &77.95  &80.16  &\textbf{96.34} &82.30 &3.54 \\

TransResUNet~\cite{tomar2022transresu} & IEEE CMBS 2023 &86.38 &78.23  &\textbf{80.85}   &95.77 & 82.83 &3.54 \\

\textbf{PVTFormer} & Proposed &\textcolor{black}{\textbf{86.78}} &\textcolor{black}{\textbf{78.46}} &80.70 & {96.11} &\textbf{82.86}	&\textbf{3.50}
\\

\bottomrule
\end{tabular}
\label{tab:segmentationlITS}
\end{table*}

\section{Results and Discussion}
Table~\ref{tab:segmentationlITS} shows the model performance of the state-of-the-art medical image segmentation architecture on Liver Tumor Segmentation Benchmark (LiTS) datasets. We have compared our proposed method with methods such as UNet~\cite{ronneberger2015u}, ResUNet++~\cite{jha2019resunet++}, DoubleU-Net~\cite{jha2020doubleu}, ColonSegNet~\cite{jha2021real}, UNext~\cite{valanarasu2022unext}, TransNetR~\cite{jha2023transnetr}, NanoNet~\cite{jha2021nanonet}, and TransResUNet~\cite{tomar2022transresu}. 
 In our rigorous evaluation, it can be observed that most of the models show strong performance for healthy liver segmentation. DoubleUNet~\cite{jha2020doubleu} that has VGG-19~\cite{simonyan2014very} as backbone notably achieved a competitive dice coefficient of 86.24\% and mIoU of 77.89\%, outperforming the competitive benchmarks such as UNext~\cite{valanarasu2022unext}, UNet\cite{ronneberger2015u}, ResUNet++\cite{jha2019resunet++}, NanoNet\cite{jha2021nanonet} and ColonSegNet~\cite{jha2021real}. However, transformer-based methods such as TransNetR~\cite{jha2023transnetr} and TransResUNet~\cite{tomar2022transresu} that used ResNet50 as the backbone show either competitive or better performance than DoubleUNet. For example, TransNetR obtained the highest precision of 96.34\%, whereas TransResUNet obtained the highest recall of 80.85\%.

 \begin{figure}[!t]
    \centering
    \includegraphics[width = 0.6\textwidth]{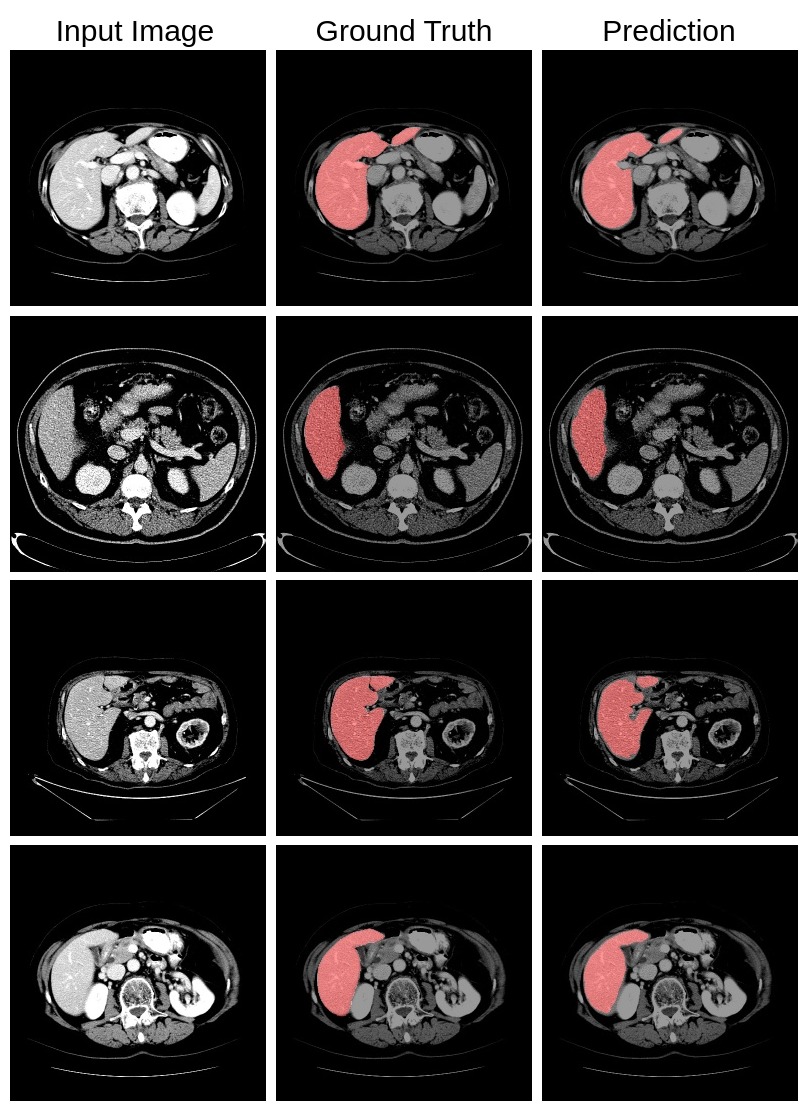}
    \caption{Components of the PVTFormer.} 
    \label{fig:qualitativeresults}
\end{figure}
 
From the table, we can also observe that the proposed PVTFormer showcased remarkable performance, demonstrating the highest dice coefficient of 86.78\%, mIoU of 78.46\%, recall of 80.70\%, precision of 96.11\%, F2 score of 82.86\%, and a low HD score of 3.50. From the overall comparison, it can be demonstrated that PVTFormer outperformed eight state-of-the-art medical image segmentation architectures. This can also be observed from the qualitative results where the proposed model successfully captures intricate details~\ref{fig:qualitativeresults}. Notably, our proposed method captures fine details and contextually significant features, surpassing the CNN-based architectures like ResUNet++\cite{jha2019resunet++}, ColonSegNet~\cite{jha2021real}, and NanoNet~\cite{jha2021nanonet} and transformer based approaches such as TransNetR~\cite{jha2023transnetr} and TransResUNet~\cite{tomar2022transresu}. While comparing the computational complexity, TransNetR operates with 10.58 GMac flops and utilizes 10.58 million parameters, whereas PVTFormer uses 43.22 GMac and utilizes 45.51 million parameters. The higher computational resource is justified by the higher performance obtained by PVTFormer compared to TransNetR and other transformer and CNN-based approaches.

\section{Conclusion}
\label{conclusion}
In this study, we proposed PVTFormer architecture by leveraging a pretrained Pyramid Vision Transformer (PVT v2) as an encoder and incorporating Up block, Decoder block and residual learning for accurate liver segmentation. The hierarchical decoding strategy in decoder blocks enhances semantic features, boosting the quality of output segmentation masks. The evaluation against eight existing state-of-the-art methods showcased that PVTFormer achieves excellent results and outperforms its competitor with a high dice coefficient of 86.78\%, a mIoU of 78.46\% and a low Hausdorff distance of 3.50. The performance of the PVTFormer demonstrates that it acts as an effective method for precise healthy liver segmentation and can also be translated to other medical applications. In the future, we plan to annotate the test dataset of LiTs with the help of radiologists from our team and perform a more comprehensive study on the multi-center dataset. Moreover, we plan to extend the capabilities of PVTFormer for segmenting multi-orans in abdominal CT scans.

\subsection*{Compliance with Ethical Standards}
This research study was conducted retrospectively using human subject data made available in open access by Bilic et al.~\cite{bilic2023liver}. Ethical approval was not required as confirmed by the license attached with the open-access data.

\subsection*{Conflicts of Interest}
 The authors have no relevant financial or non-financial interests to disclose.

\section*{Acknowledgments}
The project is supported by the NIH funding: R01-CA246704, R01-CA240639, U01-DK127384-02S1, and U01-CA268808. 

\bibliographystyle{unsrt}  
\bibliography{references}

\end{document}